\documentclass[12pt]{article}

\usepackage{graphicx,psfrag,epsf}
\usepackage{enumerate}
\usepackage{natbib}
\usepackage{mathtools}
\usepackage{booktabs}
\usepackage{color}
\usepackage[english]{babel} 
\usepackage[protrusion=true,expansion=true]{microtype} 
\usepackage{amsmath,amsfonts,amsthm}
\usepackage{amssymb}
\usepackage{chngcntr}
\usepackage[toc,page]{appendix}

\usepackage{array}
\usepackage{booktabs} 
\usepackage{natbib}
\usepackage{setspace}

\usepackage{soul}
\usepackage{multirow}
\usepackage{enumitem}
\usepackage{microtype} 
\usepackage[font = small,labelfont=bf,textfont=it]{subcaption}
\usepackage{color}
\usepackage{tabularx}
\usepackage[font = small,labelfont=bf,textfont=it]{caption} 
\usepackage{footnote}
\usepackage{algorithm}
\usepackage[noend]{algpseudocode}
\usepackage{etoolbox}

\algblock{ParFor}{EndParFor}
\algnewcommand\algorithmicparfor{\textbf{for}}
\algnewcommand\algorithmicpardo{\textbf{do\ parallel}}
\algnewcommand\algorithmicendparfor{\textbf{end\ parallel\ for}}
\algrenewtext{ParFor}[1]{\algorithmicparfor\ #1\ \algorithmicpardo}
\algrenewtext{EndParFor}{\algorithmicendparfor}

\makeatletter
\def\BState{\State\hskip-\ALG@thistlm}

\newcommand{\distas}[1]{\mathbin{\overset{#1}{\kern\z@\sim}}}%
\newcommand{\bm}[1]{\mathbf{#1}}
\newsavebox{\mybox}\newsavebox{\mysim}
\newcommand{\distras}[1]{%
  \savebox{\mybox}{\hbox{\kern3pt$\scriptstyle#1$\kern3pt}}%
  \savebox{\mysim}{\hbox{$\sim$}}%
  \mathbin{\overset{#1}{\kern\z@\resizebox{\wd\mybox}{\ht\mysim}{$\sim$}}}%
}
\newtheorem{theorem}{Theorem}

\newcommand{\be}{\begin{equation}}
\newcommand{\ee}{\end{equation}}
\newcommand{\bi}{\begin{itemize}}
\newcommand{\ei}{\end{itemize}}
\newcommand{\ben}{\begin{enumerate}}
\newcommand{\een}{\end{enumerate}}
\newcommand{\stb}{\State $\bullet$ \;}

\newcolumntype{K}[1]{>{\centering\arraybackslash}p{#1}}
\allowdisplaybreaks
\DeclareMathOperator*{\argmin}{\arg\!\min}
\makeatother

\let\oldbibliography\thebibliography
\renewcommand{\thebibliography}[1]{\oldbibliography{#1}
\setlength{\itemsep}{0pt}} 

\newcommand{\blind}{1}

\patchcmd{\footnotemark}{\stepcounter{footnote}}{\refstepcounter{footnote}}{}{}

\begin{document}

\def\spacingset#1{\renewcommand{\baselinestretch}%
{#1}\small\normalsize} \spacingset{1}

\if1\blind
{
  \title{\bf Function-on-function kriging, with applications to 3D printing of aortic tissues}
  \small
  \author{Jialei Chen\thanks{H. Milton Stewart School of Industrial and Systems Engineering, Georgia Institute of Technology} \thanks{Georgia Tech Manufacturing Institute, Georgia Institute of Technology} \thanks{Corresponding author}~, Simon Mak\thanks{Department of Statistical Science, Duke University}~, V. Roshan Joseph$^{*}$~and~Chuck Zhang$^{*}$$^{\dagger}$ \thanks{This work is supported by a U.S. National Science Foundation grant CMMI-1921646 and Piedmont Heart Institute.}\hspace{.2cm}\\
}
  \maketitle
} \fi

\if0\blind
{
  \bigskip
  \bigskip
  \bigskip
  \begin{center}
    {\LARGE\bf  Function-on-function kriging, with \\ \vspace{.3cm} applications to 3D printing of aortic tissues}
\end{center}
  \medskip
} \fi

\bigskip

\vspace{-0.5cm}
\begin{abstract}

3D-printed medical prototypes, which use synthetic metamaterials to mimic biological tissue, are becoming increasingly important in urgent surgical applications.
However, the mimicking of tissue mechanical properties via 3D-printed metamaterial can be difficult and time-consuming, due to the functional nature of both inputs (metamaterial structure) and outputs (mechanical response curve).
To deal with this, we propose a novel function-on-function kriging model for efficient emulation and tissue-mimicking optimization.
For functional inputs, a key novelty of our model is the spectral-distance (SpeD) correlation function, which captures important spectral differences between two functional inputs. Dependencies for functional outputs are then modeled via a co-kriging framework.
We further adopt shrinkage priors on both the input spectra and the output co-kriging covariance matrix, which allows the emulator to learn and incorporate important physics (e.g., dominant input frequencies, output curve properties). 
Finally, we demonstrate the effectiveness of the proposed SpeD emulator in a real-world study on mimicking human aortic tissue, and show that it can provide quicker and more accurate tissue-mimicking performance compared to existing methods in the medical literature.

\end{abstract}

\noindent%
\vfill

\newpage
\spacingset{1.45} 

\section{Introduction}

\begin{figure}
\centering
\includegraphics[width=0.8\textwidth]{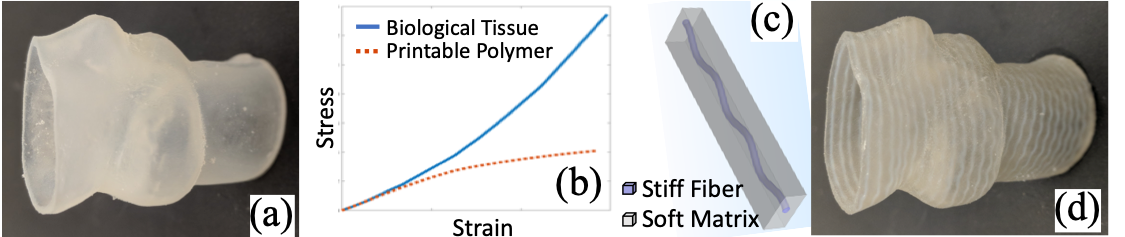}
\caption{\label{fig:3DP} (a) 3D-printed aortic valve (no metamaterial structure), (b) stress-strain curves of biological tissue and printable polymer, (c) a numerical (finite element) simulation case with sinusoidal metamaterial, (d) 3D-printed aortic valve with tissue-mimicking metamaterial.}
\end{figure}

Three dimensional (3D) printing is an emerging layer-by-layer additive manufacturing technology, with growing interest in medical applications \citep{rengier20103d}. 
This is because 3D-printed prototypes provide precise mimicking of organ shape at an acceptable price and time cost. 
Such prototypes can be extremely helpful for doctors to practice and be proficient in surgical procedures \citep{chen2018generative} as well as personalized pre-surgical planning \citep{qian2017quantitative}. 
One limitation is that the mechanical property (i.e., stress-strain curve) of printed prototypes is completely different from biological tissues \citep{raghavan1996ex}.
Currently, the state-of-the-art approach is to embed \textit{metamaterial} structure to mimic the desired mechanical property of biological tissue (\citealp{wang2016controlling}; see Figure \ref{fig:3DP}).
However, the optimization for this mimicking may take days or even weeks to perform, due to the \textit{functional} nature of the metamaterial structure.
This greatly limits the medical applicability of tissue-mimicking prototypes since surgery timing is a critical factor for outcome success.
In this paper, we propose a novel kriging model for emulating functional mechanical response over the design space of functional metamaterial structure, which can be used for efficient tissue-mimicking optimization in practical turnaround times.

There are two key reasons why state-of-the-art tissue-mimicking methods are impractical for \textit{urgent} surgical needs.
Firstly, such methods rely solely on both physical experiments and computer experiments, which are expensive and/or time-intensive to run. In particular, a single physical experiment (3D-printing and testing a prototype) takes hours to perform,
and a single computer experiment (finite element analysis)
requires at least 30 minutes for a reliable mechanical response simulation. 
Secondly, to optimize for a good structure which mimics the mechanical response of biological tissue, such methods require \textit{many} experimental runs over the design space of functional metamaterial structures. This makes current tissue-mimicking methods prohibitively expensive for urgent surgical applications, where the tissue-mimicking prototype is needed within a day. One strategy (which we adopt) is to train a \textit{surrogate} model (or \textit{emulator}, see \citealp{santner2013design}) which, given data over the design space, can efficiently \textit{predict} the mechanical response of an untested metamaterial structure. However, due to the expensive nature and functional complexities, it is necessary to integrate the rich physics of the tissue-mimicking problem within the emulator model specification, in order to achieve accurate mimicking in a timely fashion.

The proposed emulator utilizes a technique called \textit{kriging} \citep{matheron1963principles}, which models the unknown simulation output via a Gaussian process (GP).  
Kriging is widely used in computer experiment modeling for its interpolating property, and the fact that both the predictor and its uncertainty have closed-form expressions \citep{santner2013design}. 
The literature on kriging for \textit{functional outputs} typically involves some form of reduced-basis modeling \citep{bayarri2007computer,higdon2008computer,mak2018efficient,guillas2018functional}
or co-kriging framework \citep{stein1991universal,banerjee2014hierarchical}.
There has also been some work on modeling time series outputs \citep{mohammadi2019emulating}.
For \textit{functional inputs}, several techniques have been proposed in functional data analysis literature (see, e.g., \citealp{ramsay2005functional}), including varying-coefficient models \citep{fan2008statistical} and historical functional linear models \citep{malfait2003historical}.
However, the literature on kriging with functional inputs is scarce. For time-series inputs, \cite{morris2012gaussian} proposed a kriging model with a covariance function depending on time order.
Reduced-basis models were also proposed in \cite{muehlenstaedt2017computer} and \cite{wang2019gaussian}.
Such models, however, do not incorporate prior physical knowledge of the tissue-mimicking problem, and can therefore yield poor emulation 
and mimicking 
performance given the paucity and functional complexities of the experimental
data.

To address this, we introduce in this work a new function-on-function kriging model which integrates an important source of physics: the spectral information of the functional metamaterial structure input. 
Specifically, we propose a new \textit{spectral-distance} (or SpeD) correlation function, which uses the spectral-distance -- the (weighted) Euclidean distance between two functional inputs in spectral domain -- to model the process correlation of the GP. This new correlation function captures the appealing property of  \textit{translation-invariance}, where two input metamaterial structures which are the same except for a translation shift have the same mechanical properties. We then integrate this within a co-kriging framework for modeling the functional mechanical response output. This emulator-based approach allows for timely and accurate mimicking of biological tissues, and extraction of important physics (e.g., dominant input frequencies, output curve properties) via sparsity, which broadens the applicability of printed prototypes for urgent surgical procedures.

The paper is structured as follows.
Section 2 gives an overview of the tissue-mimicking problem. Section 3 presents the proposed SpeD emulation model and its shrinkage prior specification. 
Section 4 outlines the algorithm for parameter estimation. 
Section 5 investigates the emulation accuracy, uncertainty quantification, physics extraction and 
a real-world tissue-mimicking case study. 
Section 6 concludes the work.

\section{Tissue-mimicking and finite element modeling}

We first describe the tissue-mimicking problem (or the  metamaterial design
problem) and explain the physics of this problem. We then introduce the finite element (FE) analysis as a simulation tool, and provide a brief discussion on experimental design for the FE simulation.

\subsection{Tissue-mimicking problem}
As discussed, 3D-printing technology can print patient-specific prototypes with precise geometry (Figure \ref{fig:3DP} (a)), but the mechanical properties of these printed prototypes can differ greatly from that for true organs (Figure \ref{fig:3DP} (b)).
The considered mechanical property is the \textit{stress-strain} curve \citep{malvern1969introduction}, defined as stress (external tensile load per area) as a function over strain (tensile displacement as a percentage of the specimen length).
The stress-strain curve of the biological tissue typically possesses the property of \textit{strain-stiffening}, which means the curve is concave upward (see solid blue line in Figure \ref{fig:3DP} (b)), indicating it becomes stiffer as more load is introduced \citep{raghavan1996ex}. 
However, for 3D-printable material, an opposite property of \textit{strain-softening} is exhibited (see dotted red line in Figure \ref{fig:3DP} (b)) due to the plastic-slipping effect and energy dissipation \citep{hill1998mathematical}.

To achieve the strain-stiffening property of the biological tissues, one approach is to introduce \textit{metamaterial} structure (i.e., printed enhancement sub-structure) within the prototypes \citep{wang2016controlling}.
Figure \ref{fig:3DP} (c) shows an example of a metamaterial with sinusoidal structure. Here, the \textit{stiffer} enhancement fiber is designed to have a sinusoidal shape, inside the cuboid matrix of a \textit{soft} material. 
In this work, we treat the structure (or shape) of the enhancement fiber (assumed to have uniform diameter) as the functional input for our SpeD model.
Our goal is to mimic the target mechanical property of human tissues, by carefully choosing the shape of the enhancement fiber.
Figure \ref{fig:3DP} (d) shows a printed ``tissue-mimicking" aortic valve with the optimal metamaterial structure.

\subsection{Finite element modeling and experimental design}

In this work, FE modeling is used to simulate the output stress-strain curve of a given metamaterial structure.
FE modeling is frequently used for stress analysis in solid mechanics; it transforms the partial differential equations to their integral form, so that a piece-wise linear formula can be used to approximate the true deformation profile \citep{zienkiewicz1977finite}. 
The key advantage of FE simulations, compared to physical experiments, is that high accuracy can be achieved with no material cost or human error. 

Here, 
FE simulations are performed using COMSOL Multiphysics.
The overall size of the metamaterial cuboid (with one enhancement fiber inside) is $20mm$ by $4mm$ by $2mm$, with physics-based quadratic tetrahedral elements for meshing.
To compute the stress-strain curve of the metamaterial, one end of the cuboid is fixed while a series of load levels (up to 15\% uniaxial deformation) is applied to the other end. The total computation time for one metamaterial is around 30 minutes on 24 Intel Xeon E5-2650 2.20GHz processing cores.

We use a sinusoidal wave structure for designing the training metamaterial structures, as such a form exhibits the best strain-stiffening property from a recent study \citep{wang2016controlling}.
The design space has four parameters \citep{chen2018efficient}: the diameter of the enhancement fiber $d\in [0.2,2]\; mm$, and the amplitude $A \in [0,1]\; mm$, frequency $\omega\in [0,0.8]\; mm^{-1}$ and initial phase $\phi\in [0, 2\pi]$ of the sinusoidal wave:
\begin{equation}
I(t) =A \sin(2\pi\omega t + \phi).
\label{eqn:sinewave}
\end{equation}
The experimental design adopted for the sinusoidal coefficients is the maximum projection (MaxPro, \citealp{joseph2015maximum}) design,
which has good space-filling properties on design projections, thereby enabling good predictions from a GP model.
Note that the parametric sinusoidal form \eqref{eqn:sinewave} is used only to \textit{generate} data for training the emulator; we will explore a bigger non-parametric input space for \textit{prediction} and tissue-mimicking \textit{optimization}.
A total of $n=58$ metamaterial structures are simulated as the training dataset. An 18-run Sobol' sequence \citep{sobol1967distribution} is used as the testing dataset, since it provides a low-discrepancy coverage of the design space, disjoint from the training MaxPro design.
Despite the relatively small training dataset ($n=58$ samples), 
we show later that the functional stress-strain predictions from the proposed emulator are quite accurate, and provide noticeable improvements over a standard kriging model with four sinusoidal coefficients as inputs.

\section{Emulation model}

We present the proposed emulation model in three parts. 
First, we introduce the proposed model for functional inputs, using the simplified setting of scalar outputs. 
We then extend this for functional outputs using a co-kriging structure.
Finally, we discuss a prior specification for model parameters which encourages sparsity.

\subsection{Spectral-distance kriging model}

We introduce first the proposed kriging model for functional inputs $I(\cdot) \in \mathcal{I}$, where $\mathcal{I}$ is the functional input space (to be defined later). For simplicity, assume first the case of scalar outputs (functional outputs are introduced next). For the map $y(\cdot): \mathcal{I} \mapsto \mathbb{R}$ from functional inputs to scalar outputs, we propose the following GP model: 
\begin{equation}
y(\cdot) \thicksim \text{GP}\{\mu, \sigma^2 \rho(\cdot,\cdot)\},
\end{equation}
where $\mu$ is the scalar process mean and $\sigma^2$ is the process variance. Here, $\rho(\cdot,\cdot): \mathcal{I}\times \mathcal{I}\mapsto \mathbb{R}$ is the proposed spectral-distance (SpeD) correlation function, defined as:
\begin{equation}
\rho (I_1(\cdot),I_2(\cdot)) = \text{Corr}\{ y(I_1(\cdot)), y(I_2(\cdot)) \} = 
\exp \left( - \text{D}^2\left( \left|\mathcal{F}[I_1(\cdot)]\right|,\left|\mathcal{F}[I_2(\cdot)]\right|; \theta \right) \right).
\label{equ:SpeDCorr}
\end{equation}
Here, $\text{D}(\cdot,\cdot;\theta)$ is a distance function (defined later),  $|a\rm{\textbf{i}}+b|$ is the modulus of a complex number $a\rm{\textbf{i}}+b$ (where $\rm{\textbf{i}} = \sqrt{-1}$ is the unit imaginary number), and $\mathcal{F}[ \cdot ] : \mathcal{I} \rightarrow \hat{\mathcal{I}}$ is the Fourier transform from the input space of integrable functions, $\mathcal{I}=\{I(\cdot): \int  |I(t)| dt < \infty\} $, to its spectral space $\hat{\mathcal{I}}$. 
We will use the following definition of a Fourier transform for an input function $I(\cdot)$:
\begin{equation}
\hat{I}(\xi)=\mathcal{F}[I(t)]= \int I(t) e^{-2\pi \rm{\textbf{i}} t \xi} dt, \quad \xi \in \mathbb{R}.
\label{equ:FT}
\end{equation}

Similar to the scale-parametrized distance function in the Gaussian correlation (which is widely used for GP emulation of computer experiments, see \citealp{santner2013design}), we will use the following scale-parametrized $l_2$ distance function in the \textit{spectral} domain:
\begin{equation}
\text{D}(\left|\mathcal{F}[I_1(\cdot)]\right|,\left|\mathcal{F}[I_2(\cdot)]\right|; \theta )  = \left[
 \int \theta(\xi)\left(\left|\hat{I}_1(\xi)\right|-\left|\hat{I}_2(\xi)\right| \right)^2 d\xi \right]^{1/2}.
\label{equ:Dist}
\end{equation}
Here, $\theta(\cdot)$ is a weight function in spectral space, with a larger value of $\theta(\xi)$ indicating greater importance of frequency $\xi$ in the SpeD correlation function. 
In contrast to the standard Gaussian correlation, we assign importance to each \textit{frequency component} of a functional input, rather than to each input \textit{variable}. 
Plugging (\ref{equ:Dist}) into (\ref{equ:SpeDCorr}), the SpeD correlation function becomes:
\begin{equation}
\rho (I_1(\cdot),I_2(\cdot)) = 
\exp \left( - \int \theta(\xi)\left( \left|\hat{I}_1(\xi)\right|-\left|\hat{I}_2(\xi)\right|\right)^2 d\xi \right).
\label{equ:SpeDKer}
\end{equation}
In our implementation (see Section 4), this correlation is computed via a discrete approximation of the integral in \eqref{equ:SpeDKer}.

\begin{figure}
\centering
\includegraphics[width=0.67
\textwidth]{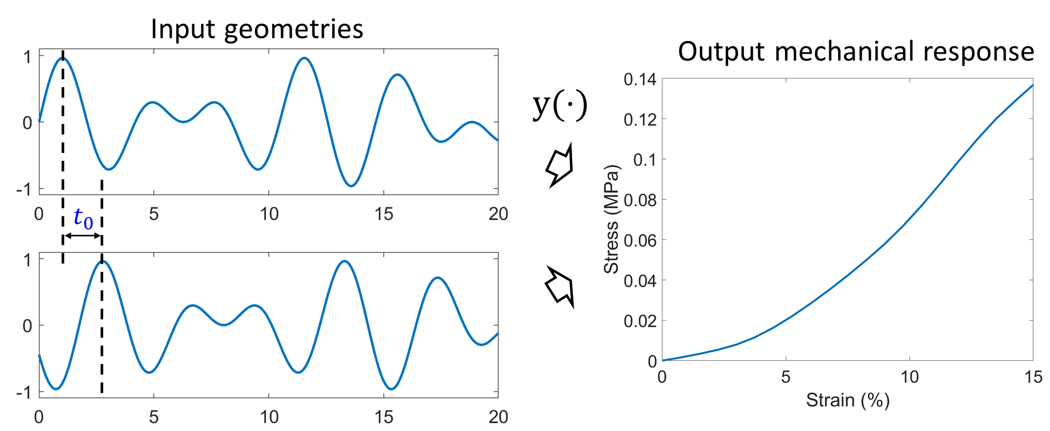}
\caption{\label{fig:TranIn} An illustration of the translation-invariance property: for the two  input structures which are equivalent up to a translation shift of $t_0$, their mechanical responses are the same.}
\end{figure}

One advantage of the SpeD correlation function is that it can capture known properties of the tissue-mimicking problem. 
First, recall that the \textit{translation-shifting} property of Fourier transform \citep{bracewell1986fourier}: for any $t_0>0$, if $I_2(t)=I_1(t-t_0)$, then
\begin{equation}
\hat{I}_2(\xi)=e^{-2\pi \rm{\textbf{i}} t_0 \xi} \hat{I}_1(\xi).
\label{equ:TIP1}
\end{equation}
For two metamaterial structures with a shift, i.e., $I_1(t)=I(t)$ and $I_2(t)=I(t-t_0)$, we can then show that their outputs are perfectly correlated, i.e.:
\begin{equation}
\rho (I_1(\cdot),I_2(\cdot))  = \exp \left(-
 \int \theta(\xi)\left(\left|\hat{I}_1(\xi)\right|-\left|e^{-2\pi \rm{\textbf{i}} t_0 \xi} \hat{I}_1(\xi)\right| \right)^2 d\xi \right) = 1.
\label{equ:TIP}
\end{equation}
We call this the \textit{translation-invariance} property of the SpeD correlation. As illustrated in Figure \ref{fig:TranIn}, this is a desirable property, since we know from physical knowledge that any translation of the metamaterial structure does not affect the output mechanical response. 
To contrast, the existing functional input models in Section 1 do not enjoy this property.
Second, it is known that the stress-strain curve depends largely on frequency $\omega$ and amplitude $A$, but not on initial phase $\phi$ in the sinusoidal parametrization \eqref{eqn:sinewave} \citep{wang2016controlling,chen2018efficient}. One can therefore expect that (i) the Fourier frequencies $\xi$ are significant, and (ii) variations in mechanical response are largely due to differences in frequency intensities $|\hat{I}(\xi)|$. The proposed correlation function (\ref{equ:SpeDKer}) nicely captures both of these properties.

For our tissue-mimicking problem, the specific choice of the Fourier transform with $l_2$ distance of the modulus gives an intuitive parametrization of known physical properties. For other applications, the SpeD correlation (\ref{equ:SpeDKer}) can also be used with other spectral transforms (e.g., wavelet transforms) and other distance metrics (e.g., $l_1$ distance). The choice of spectral transform and distance should be made on a case-by-case basis, motivated by prior information from the problem at hand.

The following theorem ensures that the SpeD correlation function $\rho (\cdot,\cdot)$ (\ref{equ:SpeDKer}) is a valid positive semi-definite kernel.

\begin{theorem}
The SpeD correlation function $\rho(\cdot,\cdot): \mathcal{I}\times \mathcal{I}\mapsto \mathbb{R}$ in (\ref{equ:SpeDKer}),
is a positive semi-definite kernel, i.e.: 
\begin{equation}
\sum_{i=1}^n \sum_{j=1}^n c_i c_j \rho (I_i(\cdot),I_j(\cdot)) \geq 0,
\label{equ:T2}
\end{equation}
holds for any $n \in \mathbb{N}, c_1, \cdots,c_n \in \mathbb{R}$ and any distinct functions $I_1(\cdot),\cdots, I_n(\cdot) \in \mathcal{I}$.
\label{Thm:PSD}
\end{theorem}
\noindent The proof of Theorem \ref{Thm:PSD} is provided in Appendix A.
This positive semi-definite property ensures the validity of $\rho(\cdot,\cdot)$ as a proper correlation function to use for GP modeling. Note that $\rho(\cdot,\cdot)$ is not (strictly) \textit{positive-definite}, in that an equality in \eqref{equ:T2} does not imply $c_i = 0$ for all $i = 1, \cdots, n$. This can be seen by setting all input functions $(I_i(\cdot))_{i=1}^n$ to be the same modulo a translation shift; the resulting correlation matrix ${[ \rho(I_i(\cdot),I_j(\cdot)) ]_{i=1}^n}_{j=1}^n$ then becomes a matrix of ones, which is clearly not positive definite. The fact that $\rho(\cdot,\cdot)$ is not positive-definite is not an issue, since for most space-filling designs  (including the adopted MaxPro design, see \citealt{joseph2015maximum}), all training input functions are distinct even after translation shifts.

\subsection{Spectral-distance co-kriging model}

For the tissue-mimicking problem, the output (i.e., the stress-strain curve) is of functional form as well. Below, we generalize the scalar model in Section 3.1 to account for functional outputs. Denote the functional input as $I(\cdot) \in \mathcal{I}$ and functional output as $O(\cdot)$, where $O(s)$ is the output stress at strain level $s$. 
For our training dataset of $n=58$ simulated structures, the functional outputs $O_i(\cdot), i =1, \cdots, n$ are discretized into $m$ levels, yielding output vectors $\bm{y}_i \in \mathbb{R}^m$, $i = 1, \cdots, n$.
We assume the following SpeD co-kriging model on $\mathbf{y}(\cdot): \mathcal{I} \mapsto \mathbb{R}^m$:
\begin{equation}
\mathbf{y}(\cdot) \thicksim \text{GP}\{\boldsymbol\mu, \boldsymbol C(\cdot,\cdot)\},
\label{equ:cokriging}
\end{equation}
where $\boldsymbol\mu \in \mathbb{R}^m$ is the process mean vector and $\boldsymbol C(\cdot,\cdot): \mathcal{I} \times \mathcal{I} \mapsto \mathbb{R}^{ m \times m }$ is the corresponding covariance matrix function.

Consider first the specification of the covariance matrix function $\boldsymbol C(\cdot,\cdot)$. Let
\begin{equation}
\boldsymbol C(I_1(\cdot),I_2(\cdot)) = \text{Cov}( \mathbf{y}(I_1(\cdot)) , \mathbf{y}(I_2(\cdot)) ) =  \rho(I_1(\cdot),I_2(\cdot))\boldsymbol\Sigma \quad \text{and} \quad \boldsymbol\Sigma \succeq 0.
\label{equ:OV}
\end{equation}
Here, $\rho(\cdot,\cdot)$ is the SpeD correlation kernel in (\ref{equ:SpeDKer}), and $\boldsymbol\Sigma \in \mathbb{R}^{m\times m}$ is a symmetric, positive definite co-kriging covariance matrix quantifying correlations between different output levels. 

Equation (\ref{equ:OV}) implicitly assumes separability in the co-kriging covariance structure. 
Here, separability means the covariance between output levels observed at different functional inputs can be decomposed as the product of the covariance between output levels and the covariance between functional inputs. This separability assumption is used extensively in the literature for reducing computational complexity \citep{banerjee2014hierarchical}. 

Consider next the specification of mean $\boldsymbol{\mu}$. We assume $\boldsymbol{\mu}$ follows the basis representation:
\begin{equation}
 \boldsymbol\mu = \mathbf{P} \boldsymbol\beta,
\label{equ:mu}
\end{equation}
where each column of $\mathbf{P} \in \mathbb{R}^{n\times q}$ represents a pre-specified basis function and $\boldsymbol\beta \in \mathbb{R}^q$ denotes its coefficients. This basis representation is similar to the modeling framework of \cite{liu2017dimension}.
The choice of basis functions in $\mathbf{P}$ should be guided by prior knowledge on the form of output stress-strain curves.  We will describe in Section 5.1 a specific parametrization of $\boldsymbol{\mu}$ which incorporates monotonicity information on the stress-strain curve.

Now, we derive the equations for prediction and UQ.
Let ${\mathbf{y}_{1:n} }= \left[ \mathbf{y}_1^T, \mathbf{y}_2^T, ..., \mathbf{y}_n^T \right]^T $ denote the vector of functional outputs of the whole training set. Using the conditional distribution formula of the multivariate normal distribution, the discretized functional response $\mathbf{y}_{\rm new}$ at a new functional input $I_{\rm new}(\cdot) \in \mathcal{I}$ follows the multivariate normal distribution:
\begin{align}
\begin{split}
\mathbf{y}_{\rm new}|{\mathbf{y}_{1:n}} \thicksim \mathcal{N} \Big( \mathbf{P}_{\rm new} \boldsymbol \beta +\left( \mathbf{r}_{\theta} \otimes \boldsymbol\Sigma \right)^T \left( \mathbf{R}_{\theta}^{-1} \otimes \boldsymbol\Sigma^{-1} \right) \left({\mathbf{y}_{1:n}}-\textbf{1}_n \otimes  \mathbf{P} \boldsymbol \beta\right), 
\\
 \boldsymbol\Sigma-\left( \mathbf{r}_{\theta} \otimes \boldsymbol\Sigma \right)^T \left( \mathbf{R}_{\theta} ^{-1} \otimes \boldsymbol\Sigma^{-1} \right) \left( \mathbf{r}_{\theta} \otimes \boldsymbol\Sigma \right) \Big),
\end{split}
  \label{equ:pred11}
\end{align}
where $\otimes$ is the Kronecker product, $\textbf{1}_n$ denotes 1-vector of $n$ elements, $\mathbf{P}_{\rm new}$ denotes the regression matrix at the new input, $\boldsymbol\beta$ and $\boldsymbol\Sigma$ are regression coefficients and co-kriging covariance matrix, $\mathbf{r}_{\theta}=\big[\rho(I_{\rm new}(\cdot), I_1(\cdot)), \cdots,$ $\rho(I_{\rm new}(\cdot),I_n(\cdot)) \big]^T$ and $\mathbf{R}_{\theta} = {[\rho(I_i(\cdot),I_j(\cdot))]_{i=1}^n} _{j=1}^n$. After algebraic manipulations, the posterior mean $\hat{\mathbf{y}}_{\rm new}=\mathbb{E}\{{\mathbf{y}}_{\rm new}|\mathbf{y}_{1:n}\}$ and posterior variance $\text{Var}\{{\mathbf{y}}_{\rm new}|\mathbf{y}_{1:n}\}$ can be written in a more concise form:
\begin{align}
\label{equ:pred21}
\hat{\mathbf{y}}_{\rm new}=\mathbb{E}\{{\mathbf{y}}_{\rm new}|\mathbf{y}_{1:n}\}= &  \mathbf{P}_{\rm new} \boldsymbol \beta+\left( \mathbf{r}_{\theta}^T \mathbf{R}_{\theta}^{-1} \otimes \mathbf{I}_m  \right) \left( \mathbf{y}_{1:n}-\textbf{1}_n \otimes   \mathbf{P} \boldsymbol \beta \right),
\\
\text{Var}\{{\mathbf{y}}_{\rm new}|\mathbf{y}_{1:n}\}= & \left(1- \mathbf{r}_{\theta}^T \mathbf{R}_{\theta}^{-1}\mathbf{r}_{\theta}  \right) \boldsymbol\Sigma, 
  \label{equ:pred22}
\end{align}
where $\mathbf{I}_m$ denotes an $m \times m$ identity matrix. Equation (\ref{equ:pred21}) can be used to predict (or emulate) the stress-strain curve for a new metamaterial structure, while Equation (\ref{equ:pred22}) can be used to construct a confidence band for quantifying the uncertainty of this prediction.

\subsection{Prior specification}
\label{sec:priors}

Finally, we provide a prior specification for the model parameters $(\theta(\cdot),\boldsymbol\Sigma,\boldsymbol{\beta})$. Consider independent priors on each parameter in $(\theta(\cdot),\boldsymbol\Sigma,\boldsymbol{\beta})$.
For the weight function $\theta(\cdot)$, we assign independent exponential priors at each frequency $\xi$, i.e.:
\begin{equation}
\theta(\xi) \distas{i.i.d.} \text{Exp}(\lambda_I),
\label{equ:priortheta}
\end{equation}
where $\lambda_I$ is a rate parameter for the exponential priors. Similar to the Bayesian LASSO \citep{park2008bayesian}, the shrinkage prior \eqref{equ:priortheta} encourages \textit{sparsity} in the maximum a posteriori estimate of $\theta(\cdot)$. This sparsity is desired for two reasons.
First, this allows us to identify dominant frequencies in metamaterial structure which influence mechanical response. 
Second, sparsity in $\theta(\cdot)$ greatly speeds up the tissue-mimicking procedure using the proposed emulator, which is paramount for efficient tissue-mimicking in urgent surgical applications. We note that, in other applications where the time budget allows for a fully Bayesian implementation (see Section 4), a spike-and-slab prior \citep{ishwaran2005spike}
could be used.

For the covariance matrix $\boldsymbol\Sigma$, we assign the following prior:
  \begin{equation}
\pi\left(\boldsymbol\Sigma\right) \propto  \exp(-\lambda_o \|\boldsymbol\Sigma^{-1}\|_1).
\label{equ:priorSigma}
\end{equation}  
Here, $\lambda_o$ is a rate parameter, and $\|\cdot\|_1$ is the element-wise $l_1$ norm. The prior \eqref{equ:priorSigma} on $\boldsymbol\Sigma$ can be viewed as a shrinkage prior which encourages sparsity on the elements of the inverse covariance matrix  $\boldsymbol{\Sigma}^{-1}$ \citep{wang2012bayesian}. This corresponds to the widely-used graphical LASSO \citep{friedman2008sparse} method for sparse covariance estimation.
For our problem, this sparsity can be used to identify important and interpretable physical couplings in the stress-strain relationship (see Section 5.3.1).

For the regression coefficients $\boldsymbol\beta$, we assign a non-informative flat prior $\pi(\boldsymbol\beta) \propto 1$, since little information is known on $\boldsymbol\beta$ prior to data in our problem. A more informative prior can be used on $\boldsymbol\beta$, if additional domain knowledge is available on the mean trend of the stress-strain curve.

\section{Parameter estimation}
In implementation, the functional inputs $I(\cdot)$ 
are also discretized to $p$ levels. Let $\boldsymbol{x}_1 = (x_1^k)_{k=0}^{p-1}$ and $\boldsymbol{x}_2 = (x_2^k)_{k=0}^{p-1}$ denote the discretized input vectors for both $I_1(\cdot)$ and $I_2(\cdot)$, respectively. The proposed SpeD kernel $\rho(I_1(\cdot),I_2(\cdot))$ in (\ref{equ:SpeDKer}) can be approximated as:
\begin{equation}
\rho (\boldsymbol{x}_1, \boldsymbol{x}_2) = 
\exp \left( - \sum_{k=0}^{(p-1)/2} \theta_k \left( \left|\hat{x}^k_1 \right|-\left|\hat{x}^k_2 \right| \right)^2  \right),
\label{equ:DSK1}
\end{equation}
where $\boldsymbol\theta = (\theta_k)_{k=0}^{(p-1)/2}$ is the discretized weight vector, and $\hat{x}^k =\sum_{l=0}^{p-1} x ^l e^{-\frac{2\pi \rm{\textbf{i}}}{p} l k}$ is the $k$-th entry of the \textit{discrete} Fourier transform $\hat{\boldsymbol x}$ for $\boldsymbol{x}$.
Note that $\hat{\boldsymbol x}$ is symmetric because $\boldsymbol{x}$ is real-valued \citep{sorensen1987real}; hence, only the first half of $\hat{\boldsymbol{x}}$ is used in (\ref{equ:DSK1}).

With this input discretization, we adopt a maximum a posteriori (MAP) approach for estimating the parameters $(\boldsymbol\beta, \boldsymbol\theta, \boldsymbol\Sigma)$. 
The main reason we prefer MAP over a fully Bayesian approach is computational efficiency, for both parameter estimation and tissue-mimicking optimization.
For parameter estimation, a fully Bayesian approach typically requires Markov chain Monte Carlo sampling (MCMC; \citealp{gelman1995bayesian}). 
Given the complexities of functional inputs and outputs, MCMC sampling can take several days, which is more time-consuming than a single computer experiment run!
Furthermore, the primary application of the proposed emulator is for tissue-mimicking optimization, which typically requires \textit{many} evaluations of the emulation predictor. Therefore, it can be \textit{very} time-consuming in a fully Bayesian implementation, since \textit{each} evaluation involves an average over all MCMC samples. In urgent surgical planning, the MAP approach (described next) offers a quicker way to survey the metamaterial design space, which enables timely tissue-mimicking optimization.

From the GP model in (\ref{equ:cokriging}) and (\ref{equ:pred11}), the MAP estimation of $(\boldsymbol\beta, \boldsymbol\theta, \boldsymbol\Sigma)$ boils down to minimizing the following penalized negative log-posterior \citep{santner2013design}:
\begin{equation}
  \begin{split}
\min_{\boldsymbol\beta, \boldsymbol\theta\geq 0, \boldsymbol\Sigma \succeq 0} l_\lambda(\boldsymbol\beta, \boldsymbol\theta, \boldsymbol\Sigma ) = &\min_{\boldsymbol\beta, \boldsymbol\theta\geq 0, \boldsymbol\Sigma \succeq 0}  \big[ n\log \det \boldsymbol\Sigma + m \log \det \mathbf{R}_{\theta} + \lambda_I \|\boldsymbol\theta\|_1+\lambda_o \|\boldsymbol\Sigma^{-1}\|_1
\\
+ &\left( \mathbf{y}_{1:n}-\textbf{1}_n \otimes \mathbf{P}\boldsymbol\beta \right)^T \left( \mathbf{R}_{\theta}^{-1} \otimes \boldsymbol\Sigma ^{-1}\right)\left(\mathbf{y}_{1:n}-\textbf{1}_n \otimes \mathbf{P}\boldsymbol\beta \right)\big].
  \end{split}
  \label{equ:post}
  \end{equation}
Here, $\mathbf{R}_{\theta}$ is the correlation matrix in (\ref{equ:pred11}) with scale parameters $\boldsymbol\theta$, and $\lambda_I$ and $\lambda_o$ are the rate parameters for the shrinkage priors in Section \ref{sec:priors}.

From a regularization perspective, the two prior terms $\lambda_I \|\boldsymbol\theta \|_1$ and $\lambda_o \|\boldsymbol\Sigma^{-1}\|_1$ in the negative log-posterior \eqref{equ:post} can equivalently be viewed as penalty terms on $\boldsymbol{\theta}$ and $\boldsymbol{\Sigma}^{-1}$,
with the rate parameters $\lambda_I$ and $\lambda_o$ corresponding to penalization parameters.
In this sense, the parameters $\lambda_I$ and $\lambda_o$ control the degree of sparsity imposed on $\boldsymbol{\theta}$ and $\boldsymbol{\Sigma}^{-1}$, with a larger $\lambda_I$ (or $\lambda_o$) resulting in a sparser estimate of $\boldsymbol{\theta}$ (or $\boldsymbol{\Sigma}^{-1}$), and vice versa. In practice, these penalization parameters can be estimated from the data itself, or specified from the problem at hand. For example, if predictive accuracy of the emulator is the end goal, then $\lambda_I$ and $\lambda_o$ can be estimated based on cross-validation techniques \citep{friedman2001elements}. However, if the extraction of important physics is desired, then $\lambda_I$ and $\lambda_o$ can be set so that a desired number of physical features can be learned. We will return to this in Section 5.3.

\begin{algorithm}[t]
\caption{BCD algorithm for minimizing the penalized negative log-likelihood (19)}\label{BCD}
\label{CD}
\begin{algorithmic}[1]
\small
\stb Set initial values $\boldsymbol{\beta} \leftarrow \bm{0}_q$, $\boldsymbol\Sigma \leftarrow \bm{I}_{m}$ and $\boldsymbol{\theta} \leftarrow \bm{1}_p$, and set ${\mathbf{Y}} \leftarrow \left[ \mathbf{y}_1, \mathbf{y}_2, ..., \mathbf{y}_n \right]^T$
\Repeat\\
\quad \quad \quad \underline{Optimizing $\boldsymbol{\Sigma}$}: 
\stb Set $\mathbf{R}_{\theta} = {\left[\exp \left( - \sum_{k=0}^{(p-1)/2} \theta_q \left( \left|\hat{x}_i^k \right|-\left|\hat{x}_j^k  \right| \right)^2  \right)\right]_{i=1}^n} _{j=1}^n$ with $\hat{x}^k =\sum_{l=0}^{p-1} x ^l e^{-\frac{2\pi \rm{\textbf{i}}}{p} l k}$
\stb Set $\boldsymbol\mu=\mathbf{P} \boldsymbol\beta$ 
\stb Set $\bm{W}_0 \leftarrow \frac{1}{n}{(\mathbf{Y}  - \bm{1}_n \otimes \boldsymbol{\mu}^T )^T \mathbf{R}_{\theta}^{-1}(\mathbf{Y} - \bm{1}_n \otimes \boldsymbol{\mu}^T )} + \lambda_o \cdot \bm{I}_{m}$
\stb Estimate $\bm{W}$ by Graphical LASSO using $\bm{W}_0$ as initialization
\stb Update $\boldsymbol\Sigma \leftarrow \bm{W}^{-1}$\\
\quad \quad \quad \underline{Optimizing $\boldsymbol\beta$}:
\stb Set $\mathbf{S} = (\mathbf{P} \otimes \bm{1}_n)^T\left(\mathbf{R}_{\theta}^{-1}\otimes\mathbf{W}\right)(\mathbf{P} \otimes \bm{1}_n)$
\stb Update $\boldsymbol\beta \leftarrow \mathbf{S}^{-1} (\mathbf{P} \otimes \bm{1}_n)^T\left(\mathbf{R}_{\theta}^{-1}\otimes\mathbf{W}\right) \mathbf{y}_{1:n}$\\
\quad \quad \quad \underline{Optimizing $\boldsymbol{\theta}$}:
\stb Update $\boldsymbol{\theta} \leftarrow \argmin_{\theta} l_\lambda(\boldsymbol{\beta}, \boldsymbol\Sigma, \boldsymbol{\theta})$ with L-BFGS 
\Until{$\boldsymbol{\beta}$, $\boldsymbol\Sigma$ and $\boldsymbol{\theta}$ converge}
\stb \Return $\boldsymbol{\beta}$, $\boldsymbol\Sigma$ and $\boldsymbol{\theta}$
\normalsize
\end{algorithmic}
\end{algorithm}

Consider now the MAP optimization in \eqref{equ:post} for fixed $\lambda_I>0$ and $\lambda_o>0$. We will use the following blockwise coordinate descent (BCD) optimization algorithm, described below. First, assign initial values for $\boldsymbol\beta$, $\boldsymbol\theta$ and $\boldsymbol\Sigma$. Next, iterate the following three steps until the convergence is achieved: (i) for fixed GP parameters $\boldsymbol\theta$ and regression coefficients $\boldsymbol\beta$, compute the correlation matrix $\mathbf{R}_{\theta}$ and then optimize for covariance matrix $\boldsymbol\Sigma$ using the graphical LASSO algorithm \citep{friedman2008sparse}; (ii) for fixed  $\boldsymbol\theta$ and $\boldsymbol\Sigma$, compute $\boldsymbol\beta$ using closed-form expressions (see \citealp{santner2013design} for details); and (iii) for fixed  $\boldsymbol\beta$ and $\boldsymbol\Sigma$, optimize for $\boldsymbol\theta$ using the L-BFGS algorithm \citep{liu1989limited}.
The full optimization procedure is provided in Algorithm 1.
Since (\ref{equ:post}) is a non-convex optimization problem, the proposed BCD algorithm only converges to a stationary solution \citep{tseng2001convergence}. Because of this, we suggest performing multiple runs of Algorithm 1 with random initializations for each run, then taking the converged estimates for the run with smallest negative log-likelihood.

\section{Emulation results}
In this section, we present the numerical performance of the proposed model for tissue-mimicking. This is presented in four parts. 
First, we compare the predictive performance of the proposed SpeD emulation model with two baseline emulation models. 
Second, we provide a comparison of the uncertainty quantification from these three emulation models.
Third, we analyze the physical properties learned via shrinkage priors on $\boldsymbol\theta$ and $\boldsymbol\Sigma$.
Finally, we demonstrate the usefulness of the fitted model for mimicking human aortic tissue.

\subsection{Prediction accuracy}
As mentioned in Section 2.2, the proposed SpeD model is fitted using the training data of $n=58$ FE simulations. 
The input function $I(\cdot)\in \mathcal{I}$ is discretized to $p=81$ parts at $\{0,0.25,0.5,\cdots, 20\}\;mm$, which we denote as a vector $\boldsymbol{x} \in \mathbb{R}^{81}$. This corresponds to the discretized $\boldsymbol\theta$ at frequencies $\{0,0.05,0.1,\cdots, 2\}\;mm^{-1}$. In the specific tissue-mimicking problem, the diameter of the metamaterial enhancement $d \in \mathbb{R}$ (assumed to be uniform over the whole functional curve $I(\cdot)$) is also important.
To account for this extra design variable, we use the following separable correlation function $\rho_s (\cdot,\cdot): \mathbb{R}^{82} \times \mathbb{R}^{82} \mapsto \mathbb{R}$:
\begin{equation}
\rho_s ([d_1,\boldsymbol{x}_1], [d_2,\boldsymbol{x}_2]) = \rho (\boldsymbol{x}_1, \boldsymbol{x}_2)
\exp \left( - \theta_d \left( d_1-d_2\right)^2  \right),
\label{equ:DSK}
\end{equation}
where $\rho(\cdot,\cdot)$ is the discretized SpeD kernel in (\ref{equ:DSK1}) and $\theta_d$ is the scale parameter for diameter $d$. 
Let $\mathbf{s}\in \mathbb{R}^{m=41}$ denote the vector of strain levels equally spaced from $s=0$ to $s=15\%$, and 
let $\mathbf{y} = O(\mathbf{s}) \in \mathbb{R}^{41}$ be the discretized stress function $O(\cdot)$. Here, the input and output discretization levels are selected heuristically to capture features of the input and output functions: the output functions are quite smooth require less levels, and the input functions are more rugged and require more levels.

From the underlying physics of the stress-strain relationship, it is known that (i) the stress $O(s)$ is always positive, (ii) the stress is zero when the strain is zero (this is known as the free-standing state, see \citealp{malvern1969introduction}), and (iii) stress-strain curves are typically monotone and non-decreasing, since a larger force is needed to stretch further. To account for (i), a standard log-transformation of stress $O(s)$ is performed prior to modeling and parameter estimation, and the final results are transformed back to ensure the predicted stress is always positive. To account for (ii) and (iii), we choose the basis functions in \eqref{equ:mu} to be $\mathbf{P}=[\textbf{1}_m,\log(\mathbf{s})]$, along with an additional constraint of $\beta_2>0$ to ensure the mean function is monotone and non-decreasing.  This is equivalent to assuming the mean stress-strain curve takes the following form $O(s)=a s^{b},\;a,b>0$, which is a typical parametrization in biomedical literature \citep{rengier20103d,chen2018efficient}. 
This provides a simple and effective way to encourage monotonicity via the mean function specification; one can also extend the shape-constrained GP model in \cite{wang2016estimating} to impose sample path monotonicity, but this is beyond the scope of this work.

\begin{figure}
\centering
\includegraphics[width=0.8
\textwidth]{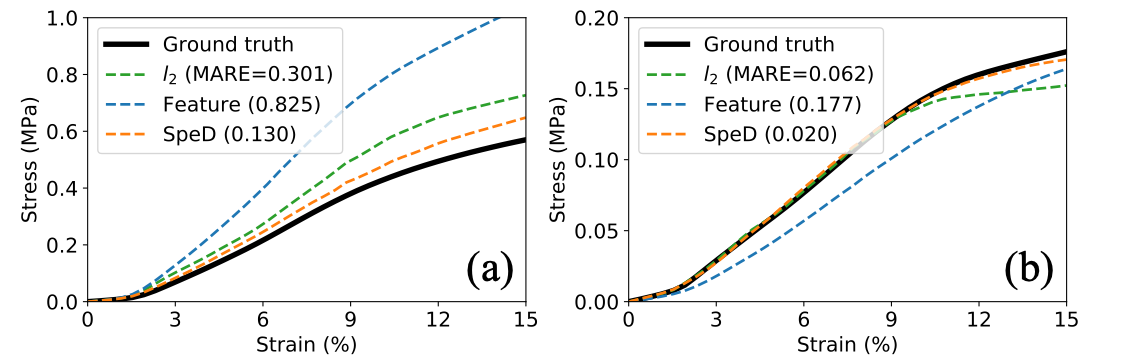}
\caption{\label{fig:comp} Predicted stress-strain curves for the $l_2$-distance emulator (``$l_2$''), feature-based emulator (``Feature''), and the proposed SpeD emulator (``SpeD'') on two test metamaterial structures. The corresponding MAREs are included in the legends.} 
\end{figure}

For comparison, we also fit two different emulators as baseline methods, using the same dataset. 
The inputs of the first emulator are the parameters from the sinusoidal wave design $\boldsymbol{x}_p=[d,A,\omega,\phi]^T\in \mathbb{R}^4$, which represents the diameter of the metamaterial fiber, amplitude, period and initial phase of the sinusoidal wave (see Figure \ref{fig:3DP} and Equation \eqref{eqn:sinewave}). This emulator uses a GP model with correlation function:
\begin{equation}
\rho_p (\boldsymbol{x}_{1}, \boldsymbol{x}_{2}) = 
\exp \left( - \sum_{k=1}^4 \theta_k \left( {x}_{1}^k -{x}_{2}^k\right)^2  \right).
\label{equ:PK}
\end{equation}
The same correlation function (with scalar output) is used in \cite{chen2018efficient}.
We refer this as the \textit{feature-based} method. The second emulator also assumes a GP model with correlation function:
\begin{equation}
\rho_{f} (I_1(\cdot), I_2(\cdot)) = 
\exp \left( - \int \theta(t) \left( I_1(t) -I_2(t)\right)^2 dt \right).
\label{equ:DFK}
\end{equation}
This correlation (\ref{equ:DFK}) is essentially the Gaussian correlation function, with distance taken to be the $l_2$-distance between input functions. A similar correlation function is used in \cite{morris2012gaussian} for time-series inputs, with additional dependencies on time order.
We refer this as the functional \textit{$l_2$-distance} method. Both baseline methods assume the same separable co-kriging structure for discretized outputs $\mathbf{y}_{1:n}$, along with MAP parameter estimation.

\subsubsection{Predicting stress-strain curve}

\begin{table}[t]
\begin{minipage}{0.6\textwidth}
\centering
\includegraphics[width=0.65\textwidth]{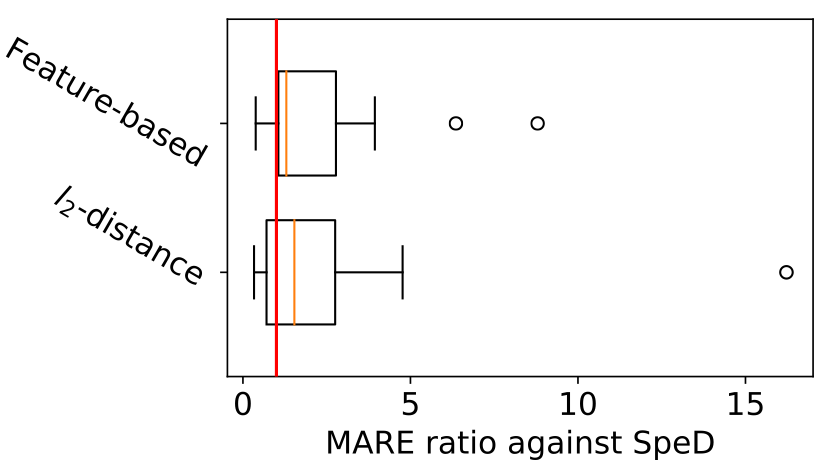}
\captionof{figure}{\label{fig:hist} Boxplots of the MARE ratio between the baseline emulators and SpeD emulator on the 18-run test set. The red line marks the MARE ratio of 1.0, where the baseline emulator has the same MARE as the SpeD emulator.}
\end{minipage}
\hfill
\begin{minipage}{0.38\textwidth}
\begin{tabular}{c c} 
 \toprule
  & \textbf{Median MARE}\\ 
 \toprule
SpeD & \textbf{0.11}   \\
Feature-based & 0.19    \\
$l_2$-distance &  0.26  \\
 \toprule
\end{tabular}
\caption{The median MARE of the SpeD emulator and two baseline emulators over the 18-run test set.}
\label{tab:MARE}
\end{minipage}

\end{table}

To test the performance of the proposed emulator, we compare the predictions of stress-strain curves (using Equation (\ref{equ:pred21})) for the metamaterial designs from the test set (see Section 2.2).
Figure \ref{fig:comp} shows the emulated stress-strain curves for two test metamaterial structures, along with the true stress-strain curve (ground truth) from FE simulations.
To quantitatively measure the difference between the predicted and true curves, we use the following mean absolute relative error (MARE) metric:
\begin{equation}
\text{MARE}=\frac{\int_s |O(s)-\hat{O}(s)|ds}{\int_s |O(s)|ds},
\label{equ:MARE}
\end{equation}
where $s$ is the strain level, $O(s)$ is the stress at strain $s$ from FE simulation (ground truth), and $\hat{O}(s)$ is the predicted stress from the emulators. 
The MARE values for the two test cases in Figure \ref{fig:comp} are included in the legends. 
Table \ref{tab:MARE} reports the median MARE values for the three considered emulators, over the whole test set. 
The proposed SpeD emulator appears to perform very well, in that it achieves noticeably lower median MARE than the two existing emulators. 
Figure \ref{fig:hist} shows the boxplots of the MARE ratio between the baseline emulators and the SpeD emulator for the 18 test cases (note that a ratio of 1.0 means the SpeD model yields similar MARE to a baseline model).
We see that these ratios are mostly larger than one, which suggests that the proposed emulator is noticeably better in predicting the true stress-strain output curve. This is not surprising, since our model captures known physical properties of the tissue-mimicking problem.

\subsubsection{Predicting physical characteristics}
 \begin{figure}[t]
\centering
\includegraphics[width=0.95\textwidth]{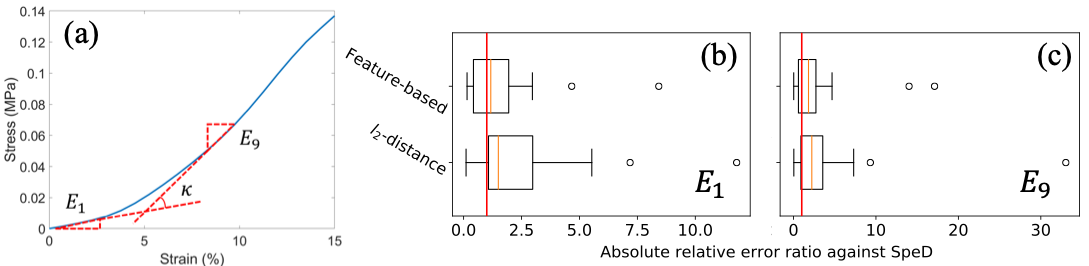}
\caption{\label{fig:chara} (a) visualizes the three characteristics of mechanical performance: moduli $E_1$ and $E_9$, and curvature $\kappa$. 
(b) and (c) show the pairwise absolute relative error for $E_1$ and $E_9$ between the two baseline emulators and the SpeD emulator. The red line marks a relative error ratio of 1.0.}
\end{figure}

In addition to predicting stress-strain curve $O(s)$, engineers are also interested in predicting key physical characteristics.
An accurate prediction of these characteristics can be as important as emulating the stress-strain curve itself, because it provides interpretability to the black-box emulation model. 
Two important physical characteristics of interest are (i) the elastic modulus of the stress-strain curve, and (ii) the classification of material type as strain-stiffening or strain-softening. For (i), the \textit{modulus}, i.e., the slope of the stress-strain curve at different strain levels, can be interpreted as the stiffness or hardness of the material \citep{raghavan1996ex}.
Here, we are interested in the elastic moduli $E_1$ and $E_9$ at strain levels $1\%$ and $9\%$, respectively, where
$ E_k=\partial O(s)/ \partial s\big|_{s=k\%}$; this allows us to evaluate the elastic moduli prediction over a wide range of strain levels.
For (ii), we wish to classify the stress-strain curve as \textit{strain-stiffening} or \textit{strain-softening}; this is particularly important given the goal of mimicking biological tissues (see Section 2.1). 
One way to classify is to use the curvature of the stress-strain curve, which can be approximated by the slope of the two moduli,
$\kappa=\partial^2 O / \partial s^2 \approx (E_9-E_1)/(9\%-1\%)$.
Assuming no fluctuations in $s \in [1,9]\%$ \citep{malvern1969introduction}, a positive curvature $\kappa$ 
suggests a strain-stiffening property is present (due to increasing moduli), while a negative $\kappa$ suggests a strain-softening property is present.
Figure \ref{fig:chara} (a) visualizes these physical characteristics from a stress-strain curve.

\begin{table}[t]
\centering
\begin{tabular}{c c c c} 
 \toprule
 & \textbf{SpeD} & \textbf{Feature-based}& \textbf{$l_2$-distance} \\ 
 \toprule
\textit{True positive \%} & 12/12=100\% & 11/12=91.7\% & 7/12=58.3\%  \\
\textit{True negative \%} & 6/6=100\%   & 5/6=83.3\%   & 5/6=83.3\%   \\ \hline
\textit{Classification \%}      & 18/18=100\% & 16/18=88.9\% & 13/18=72.2\% \\
 \toprule
\end{tabular}
\caption{The true positive rate, true negative rate, and classification rate
of strain-stiffening and strain-softening, for the three considered emulators.}
\label{table:ac}
\end{table}

We now compare three emulators (SpeD and baselines) for predicting the moduli and material type. The moduli $\hat{E}_1$ and $\hat{E}_9$, computed from the emulated stress-strain curves, are compared with the moduli ${E}_1$ and ${E}_9$ from FE simulation. Figures \ref{fig:chara} (b) and (c) show the pairwise absolute relative error $|\hat{E}-E|/E$, between the baselines and the SpeD emulator. We see that most of these ratios are larger than 1.0 in the test set, which shows that the proposed SpeD model outperforms both baseline emulators.
For classification, 
the predicted curvature $\hat{\kappa}$, computed from the emulated curves, are compared with the true curvature $\kappa$ from FE simulation. 
Table \ref{table:ac} shows the correct classification rates for the three emulators.
The SpeD model has a perfect $18/18=100\%$ classification accuracy: it identified the correct strain-softening/-stiffening property for all 18 test structures. On the other hand, the feature-based model and the $l_2$-distance model achieves only a $16/18=88.9\%$ and $13/18=72.2\%$  classification accuracy rate, respectively. One reason why the proposed SpeD model can better capture these physical characteristics (compared to existing emulators) 
is because it directly incorporates the underlying physics via the SpeD correlation function.

\subsection{Uncertainty quantification}

\begin{figure}
\centering
\includegraphics[width=0.92\textwidth]{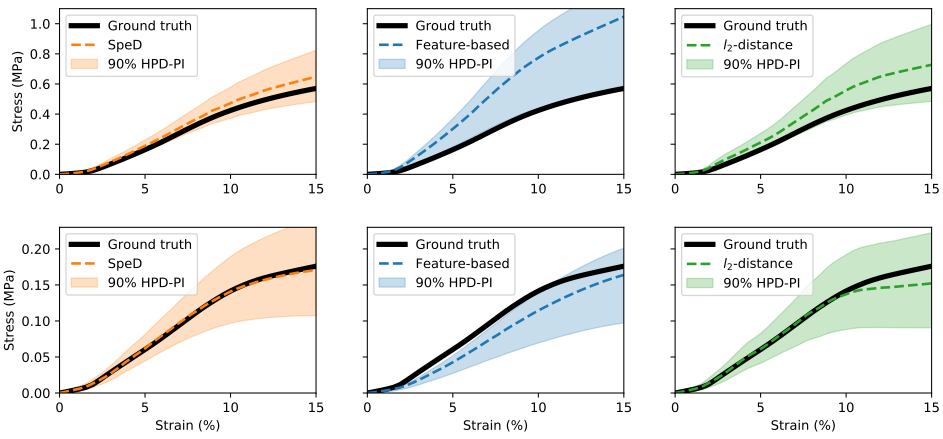}
\caption{\label{fig:UQ} A comparison of the 90\% pointwise HPD-PIs for the three emulation models (left: SpeD, middle: feature-based, right: $l_2$-distance). Different rows are for different test cases.}
\end{figure}

Particularly in healthcare applications, the quantification of predictive uncertainty can be as important as the prediction itself.
For the proposed model, Equations \eqref{equ:pred21} and \eqref{equ:pred22} can be used to construct 90\% pointwise highest posterior density predictive intervals (HPD-PIs) for the emulated stress-strain curves.
Figure \ref{fig:UQ} shows the $90\%$ HPD-PI for the three emulation models. 
Note that there is little predictive uncertainty at low strain, with uncertainty increasing as strain levels increase. This is consistent with the physical intuition in Section 5.1: the stress always equals to zero when strain equals zero, i.e., no force at free-standing condition. The increasing uncertainty for higher strain levels may be due to the log-transformation of the functional output.

Comparing the predictive intervals for the three emulators, we see that the proposed SpeD model returns narrower predictive intervals compared to both the $l_2$-distance model and the feature-based model. This is particular evident for the test case in the top row of Figure \ref{fig:UQ}. Moreover, the $90\%$ HPD-PIs of the SpeD emulator covers the true stress-strain curves in $16/18$ of the test cases, whereas the coverage for the feature-based and $l_2$-distance emulators are only $12/18$ and $14/18$, respectively. For example, the bottom row of Figure \ref{fig:UQ} shows a test case where feature-based emulator fails to cover the true stress-strain curve.
Over the whole test set, our SpeD emulator appears to give reliable coverage of the true stress-strain curve, with relatively low predictive uncertainty. The reasons for this may be two-fold: (i) the SpeD correlation captures the physics of the tissue-mimicking problem, which can be viewed as an additional source of data, and (ii) the shrinkage priors on spectral coefficients screens out inert frequencies, which also helps reduce predictive uncertainty. It is worth noting that the predictive intervals here do not account for parameter uncertainties in the emulator; accounting for such uncertainties would require a fully Bayesian implementation, which would entail much more computational resources.

\subsection{Learning physics via sparsity}

The SpeD emulator also provides a data-driven approach to learn important physics, via the shrinkage priors on both the covariance matrix $\boldsymbol\Sigma$ and frequency coefficients $\boldsymbol\theta$.

\subsubsection{Segmentation of stress-strain curve}

\begin{figure}
\centering
\includegraphics[width=0.72\textwidth]{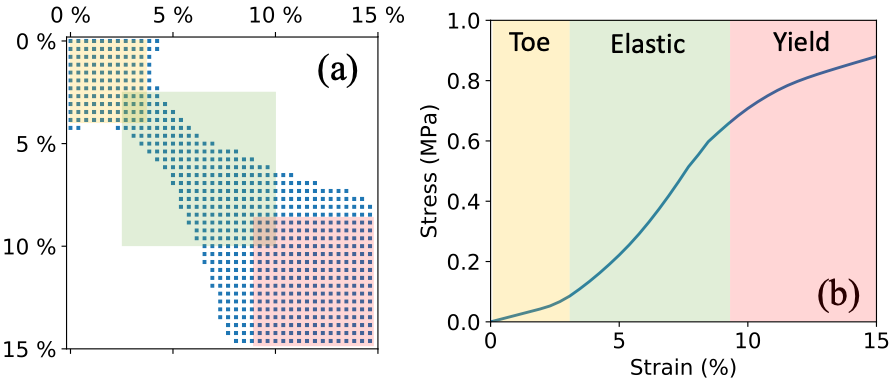}
\caption{\label{fig:spars} (a) The sparsity pattern visualization of the inverse covariance matrix $\boldsymbol\Sigma^{-1}$ by graphical LASSO with $40\%$ of non-zero entries. The pattern indicates three different regions of the stress-strain curve, colored yellow, green and red. (b) 
The partition of strain-stress curves for soft materials into toe, elastic, and yield regions up to strain level of $15 \%$.
}
\end{figure}

We first analyze the important correlations selected by the shrinkage prior on the co-kriging covariance matrix $\boldsymbol \Sigma$.
Setting the penalty parameter $\lambda_o$ such that 40\% of the entries of $\boldsymbol\Sigma^{-1}$ are non-zero, Figure \ref{fig:spars} (a) visualizes the selected (important) covariances in $\boldsymbol\Sigma^{-1}$. 
Each entry of $\boldsymbol\Sigma^{-1}$ represents the corresponding covariance between two stress-strain curve points conditional on all other curve points; note that this covariance quantifies the deviation of the curve from the parametric model $O(s)=as^b$.
We see that the stress-strain curve can be roughly segmented into three regions: small strain (from $0\%$ to $3\%$) with high conditional correlation, medium strain (from $3\%$ to $9\% $) with moderate conditional correlation and large strain (from $9\%$ to $15\% $) with high conditional correlation.

These three regions suggest a connection to known physical properties in material strength \citep{malvern1969introduction, martin1998skeletal}, where the mechanical response of the soft bio-mimicking material can also be divided to three regions: the toe region, the elastic region and the yield region (see Figure \ref{fig:spars} (b)).
We see from Figure \ref{fig:spars} (a) that there are fewer significant conditional correlations in the elastic region compared to the other two regions.
One reason for this is that, within the elastic region, the stress-strain curve can be better approximated by the form $O(s)=as^b$ (which corresponds to the choice of basis functions in $\mathbf{P}$).
Figure \ref{fig:spars} (a) also suggests the presence of conditional correlations between the elastic and yield regions. 
One plausible explanation of this is the migration of strain-stiffening or strain-softening property to straightening.

\subsubsection{Learning dominant frequencies}

\begin{figure}
\centering
\includegraphics[width=0.66\textwidth]{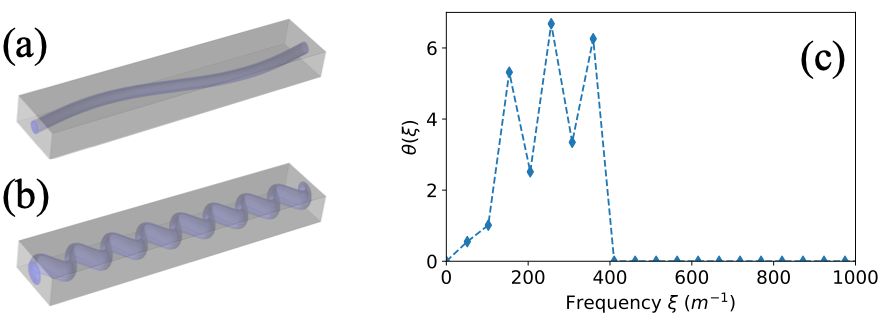}
\caption{\label{fig:coef} 
Examples of metamaterial structure with very low (a) or very high (b) frequency. 
(c) MAP estimates of spectral parameters $\theta$, where medium frequencies are non-zero.}
\end{figure}

The proposed approach can also learn important frequencies $\xi$ which influence mechanical response, via the shrinkage priors on the weight function $\theta(\xi)$  (see Section 3.3). 
Figure \ref{fig:coef} (c) shows the MAP estimate of $\theta(\cdot)$ in the spectral space, where the rate parameter $\lambda_I$ is chosen via cross-validation. We see that $\theta(\cdot)$ shrinks to zero at low and high frequencies, with non-zero estimates only for medium frequencies between $50 m^{-1}$ to $400 m^{-1}$. For these two endpoint frequencies, Figures \ref{fig:coef} (a) and (b) show the metamaterial structures with frequencies $\xi \approx 50 m^{-1}$ and $\xi \approx 400 m^{-1}$, respectively.

The selected frequencies in $\theta(\xi)$ are also in line with the physical understanding of the problem.
For low frequencies (Figure \ref{fig:coef} (a)), 
the fluctuation in metamaterial design is too weak to induce any effect on the stress-strain curve, whereas for high frequencies (Figure \ref{fig:coef} (b)), 
the resulting strong fluctuation in metamaterial leads to mechanical properties similar to a straight fiber (given nonzero diameter $d$, see \citealp{chen2018efficient}).
While it is known that different frequencies affect mechanical response in different ways, a strict law is difficult to find for engineers. 
Here, our SpeD emulator sheds light on the influential frequencies, i.e., from $50 m^{-1}$ to $400 m^{-1}$, so those frequencies should be carefully chosen for metamaterial design. We note that these selected frequencies may be sensitive to the choice of experimental design, so further analyses should be taken to confirm such findings from a physics perspective.
This identification of important frequencies also allows us to greatly speed up optimization for tissue-mimicking, which we show next.

\subsection{Mimicking aortic tissue via  optimization}

We now tackle the motivating task of mimicking the mechanical properties of a target tissue with the proposed emulator.
Here, the SpeD model can be used to find a good metamaterial design (both structure $I(\cdot)$ and diameter $d$) whose stress-strain curve matches the desired mechanical property $\mathbf{y}^*$.
This is achieved via the following optimization problem:
\begin{equation}
\left(d^*, I^*(\cdot)\right)= \argmin_{d_{\rm new},I_{\rm new}(\cdot) \in \mathcal{I}}\mathbb{E}\big\{ \| \mathbf{y}([d_{\rm new},I_{\rm new}(\cdot)])-\mathbf{y}^* \|_2^2|\mathbf{y}_{1:n}\big\},
\label{Equ:Mimic_1}
\end{equation}
where $I^*(\cdot)$ is the optimal metamaterial structure, $d^*$ is the optimal fiber diameter, and $\mathbf{y}([d_{\rm new},I_{\rm new}(\cdot)])|\mathbf{y}_{1:n}$ is the conditional (discretized) stress-strain curve in (\ref{equ:pred11}) with diameter $d_{\rm new}$ and structure $I_{\rm new}(\cdot)$.
In words, equation (\ref{Equ:Mimic_1}) aims to find the optimal metamaterial design whose stress-strain curve from the proposed model (conditional on data) is closest to the target curve $\mathbf{y}^*$ in terms of mean-squared error (MSE). 

This MSE criterion can be further decomposed as follows:
\begin{equation}
\| \hat{\mathbf{y}}([d_{\rm new},I_{\rm new}(\cdot)])-\mathbf{y}^*\|_2^2 + \textrm{tr}\left( \text{Var}\{ \mathbf{y}([d_{\rm new},I_{\rm new}(\cdot)])|\mathbf{y}_{1:n} \} \right).
\label{Equ:Mimic3}
\end{equation}
Here,
$\hat{\mathbf{y}}([d_{\rm new},I_{\rm new}(\cdot)])$ and  $\text{Var}\{ \mathbf{y}([d_{\rm new},I_{\rm new}(\cdot)])|\mathbf{y}_{1:n} \}$ are the conditional mean and variance of $\mathbf{y}([d_{\rm new},I_{\rm new}(\cdot)])|\mathbf{y}_{1:n}$, respectively, and $\textrm{tr}(\mathbf{A})=\sum_i A_{i,i}$ is the trace of the matrix $\mathbf{A}$. 
The first term can be interpreted as trying to minimize the $l_2$-norm between the emulated stress-strain curve and the target curve. The second term can be viewed as trying to minimize the predictive variance of the emulated curve. Such a decomposition is quite intuitive, since we wish to find a metamaterial design whose emulated curve matches the desired curve, but also has low predictive uncertainty from the emulation model.

One difficulty in solving (\ref{Equ:Mimic_1}) is that the variable $I(\cdot)$ is \textit{functional} in form, and its discretization $\boldsymbol x \in \mathbb{R}^p$, $p=81$ can be too \textit{high dimensional} to optimize numerically. 
Here is where the extracted important frequencies from Section 5.3.2 come into play. Let $\hat{\boldsymbol{x}}_{\rm new} \in \mathbb{R}^7$ denote the seven non-zero Fourier coefficients (see Figure \ref{fig:coef}). Using these coefficients as inputs for optimization (and ignoring the other inert coefficients), we get the following lower-dimensional optimization problem:
\begin{equation}
\left( d^*, \hat{\boldsymbol x}^*\right) = \argmin_{d_{\rm new}\in \mathbb{R},\hat{\boldsymbol x}_{\rm new}\in \mathbb{R}^{7}} \mathbb{E}\big\{ \| \mathbf{y}([d_{\rm new},\hat{\boldsymbol x}_{\rm new}])-\mathbf{y}^* \|_2^2 |\mathbf{y}_{1:n}\big\},
\label{Equ:Mimic_2}
\end{equation}
where $\mathbf{y}([d_{\rm new},\hat{\boldsymbol x}_{\rm new}])|\mathbf{y}_{1:n}$ is the conditional random vector taking the frequencies as input. While this problem is non-convex, it is much \textit{lower-dimensional}, and can be effectively optimized using standard quasi-Newton methods (e.g., L-BFGS) and random initializations.

This framework (using the proposed SpeD model) offers significant speeds up for tissue-mimicking over the current state-of-the-art methods. To see why, consider first the optimization of (\ref{Equ:Mimic_1}) using only numerical FE simulations: this requires hundreds of evaluations of the optimization objective function, each of which requires around 30 minutes of computation time. This means tissue-mimicking with only FE simulations can require many days of computation, which is clearly unsuitable for urgent surgical planning \citep{chen2018efficient}. To contrast, each evaluation of the proposed emulator requires only seconds of computation, which greatly speeds up the mimicking process. Furthermore, by exploiting sparsity in spectral coefficients, the dimension of the optimization problem reduces from 
82 to 8 variables. This dimension reduction greatly cuts down on the number of predictions from the emulator, which yields significant reductions in computation time. 
Such speed-ups are paramount for performing tissue-mimicking in an accurate and timely manner.
Section 5.5 provides a further comparison of timing.

\begin{figure}
\centering
\includegraphics[width=0.7\textwidth]{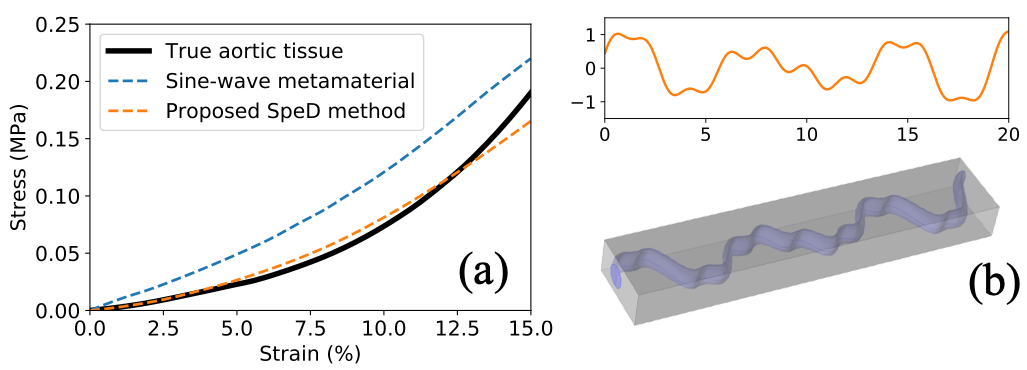}
\caption{\label{fig:aor} A case study which uses the proposed SpeD model for mimicking human aortic tissue. (a) shows the stress-strain curves of the target aortic tissue (black), the mimicked curve from an existing method (blue) and the curve optimized from the SpeD method (red). (b) shows the optimal metamaterial design from our approach.}
\end{figure}

Figure \ref{fig:aor} (a) shows the 
stress-strain curve of a target aortic tissue (in black) from \cite{hockaday2012rapid}, the stress-strain curve from the proposed mimicking procedure (in red), and the curve from an existing mimicking method (in blue) in \cite{wang2016controlling}. The latter method performs mimicking using only the four sinusoidal metamaterial features (see Section 2.2).
Compared to the existing approach,
which has an MARE (see Equation \ref{equ:MARE}) of $0.528$, the proposed SpeD approach achieves a much smaller MARE of $0.089$.
This improved tissue-mimicking performance can be seen in Figure \ref{fig:aor} (a): the red curve (from the proposed method) closely mimics the desired black curve, whereas the blue curve (from \citealp{wang2016controlling}) overestimates stress at all strain levels. In particular, our method gives much better mimicking in small strain regions -- this is important in medical applications due to the relatively small strain deformations in the human body. 

There are two reasons for this improved performance. First, the existing mimicking approach in \cite{wang2016controlling} is too restrictive, in that it uses only four sinusoidal features and not the full functional form of the input. Second, given a fixed timeframe, the proposed emulation-based approach permits a larger number of objective evaluations via the proposed SpeD model. Figure \ref{fig:aor} (b) shows the optimal (discretized) metamaterial design ${\boldsymbol x}^*$ from our emulation-based approach, which is clearly not a sinusoidal function. By considering the broader class of functional inputs as well as allowing for more objective evaluations, the proposed method can identify  better metamaterial designs for tissue-mimicking.

\subsection{Computation time}

\begin{table}[t]
\centering
\begin{tabular}{c c} 
 \toprule
\textbf{Modeling Step} & \textbf{Computation Time (in minutes)}  \\
 \toprule
 Parameter estimation &21.72 \\
Prediction \& UQ at one input setting & 0.01 \\
Tissue-mimicking of a target material & 2.69 \\
 \toprule
\end{tabular}
\caption{Computation time for different modeling steps of the proposed emulator, parallelized over 24 processing cores.}
\label{table:time}
\end{table}

Another appeal of the SpeD emulator is its computational efficiency. Table \ref{table:time} summarizes the computation time required for each step of the emulation process, with timing performed on a parallelized system of 24 Intel Xeon E5-2650 2.20GHz processing cores. The computation time required for parameter estimation (with cross-validation on $\lambda_I$) is $21.72$ minutes, which is typically performed before the arrival of the patient. Once the model is fit, we can predict for multiple settings very quickly ($0.01$ minutes for each structure). To contrast, FE simulations require 30 minutes for each structure, and a fully Bayesian implementation of the emulator, which averages over a large amount of MCMC samples (say, 2000), takes at least $0.01\times 2000 = 20$ minutes per structure.
Because of this, our SpeD emulator can effectively perform the tissue-mimicking procedure using only 3 minutes of computation; this greatly improves upon the standard tissue-mimicking approach with only FE simulations, which may require hours or even days to perform \citep{wang2016controlling} with much poorer mimicking performance (see Figure \ref{fig:aor})!
Therefore, the proposed SpeD model can provide effective and personalized pre-surgical practicing and planning \citep{qian2017quantitative,chen2019avp} with dramatically lower costs, which then mitigates risk in complex surgical procedures.

\section{Conclusion}

We propose in this paper a novel function-on-function Gaussian process emulation model for tackling the challenging tissue-mimicking optimization, under urgent surgical demands. 
The key challenge is the \textit{functional} input metamaterial structures and the \textit{functional} output mechanical responses. 
To address this, the proposed co-kriging model uses a new spectral-distance (SpeD) correlation function, which integrates spectral information by directly modeling the effect of metamaterial frequencies on mechanical response. One appealing feature of this new correlation function is its \textit{translation-invariance} property, which accounts for the fact that two metamaterial structures, which are equivalent modulo a translation shift, have the same mechanical properties.
For parameter estimation, we use MAP with shrinkage priors, which identifies key frequencies and thereby reduces the large \textit{functional} input space.
This reduction greatly speeds up the tissue-mimicking optimization using the proposed emulator.
Applied to a real-world tissue-mimicking study, the proposed SpeD emulator outperforms existing models in (i) emulating and quantifying uncertainty on mechanical response, (ii) extracting meaningful physical insights, and (iii) providing efficient and accurate mimicking performance for human aortic tissue.
One direction for future work is the exploration of a more elaborate design method for functional inputs, which may further improve emulation performance. With the development of multi-material 3D-printing technology, this new emulator can play an important role in furthering the impact of 3D-printing in important biomedical applications in surgery planning and healthcare.

\small

\begin{thebibliography}{}

\bibitem[Banerjee et~al., 2014]{banerjee2014hierarchical}
Banerjee, S., Carlin, B.~P., and Gelfand, A.~E. (2014).
\newblock {\em Hierarchical Modeling and Analysis for Spatial Data}.
\newblock CRC Press.

\bibitem[Bayarri et~al., 2007]{bayarri2007computer}
Bayarri, M., Berger, J., Cafeo, J., Garcia-Donato, G., Liu, F., Palomo, J.,
  Parthasarathy, R., Paulo, R., Sacks, J., and Walsh, D. (2007).
\newblock Computer model validation with functional output.
\newblock {\em The Annals of Statistics}, 35(5):1874--1906.

\bibitem[Bracewell and Bracewell, 1986]{bracewell1986fourier}
Bracewell, R.~N. and Bracewell, R.~N. (1986).
\newblock {\em The {Fourier} Transform and its Applications}, volume 31999.
\newblock McGraw-Hill New York.

\bibitem[Chen et~al., 2018a]{chen2018efficient}
Chen, J., Wang, K., Zhang, C., and Wang, B. (2018a).
\newblock An efficient statistical approach to design {3D}-printed
  metamaterials for mimicking mechanical properties of soft biological tissues.
\newblock {\em Additive Manufacturing}, 24:341--352.

\bibitem[Chen et~al., 2018b]{chen2018generative}
Chen, J., Xie, Y., Wang, K., Wang, Z.~H., Lahoti, G., Zhang, C., Vannan, M.~A.,
  Wang, B., and Qian, Z. (2018b).
\newblock Generative invertible networks ({GIN}): Pathophysiology-interpretable
  feature mapping and virtual patient generation.
\newblock In {\em International Conference on Medical Image Computing and
  Computer-Assisted Intervention}, pages 537--545. Springer.

\bibitem[Chen et~al., 2020]{chen2019avp}
Chen, J., Xie, Y., Wang, K., Zhang, C., Vannan, M.~A., Wang, B., and Qian, Z.
  (2020).
\newblock Active image synthesis for efficient labeling.
\newblock {\em IEEE Transactions on Pattern Analysis and Machine Intelligence}, to appear.

\bibitem[Fan and Zhang, 2008]{fan2008statistical}
Fan, J. and Zhang, W. (2008).
\newblock Statistical methods with varying coefficient models.
\newblock {\em Statistics and its Interface}, 1(1):179.

\bibitem[Friedman et~al., 2001]{friedman2001elements}
Friedman, J., Hastie, T., and Tibshirani, R. (2001).
\newblock {\em The Elements of Statistical Learning}, volume~1.
\newblock Springer Series in Statistics, NY.

\bibitem[Friedman et~al., 2008]{friedman2008sparse}
Friedman, J., Hastie, T., and Tibshirani, R. (2008).
\newblock Sparse inverse covariance estimation with the graphical lasso.
\newblock {\em Biostatistics}, 9(3):432--441.

\bibitem[Gelman et~al., 1995]{gelman1995bayesian}
Gelman, A., Carlin, J.~B., Stern, H.~S., and Rubin, D.~B. (1995).
\newblock {\em Bayesian Data Analysis}.
\newblock Chapman and Hall/CRC.

\bibitem[Guillas et~al., 2018]{guillas2018functional}
Guillas, S., Sarri, A., Day, S.~J., Liu, X., and Dias, F. (2018).
\newblock Functional emulation of high resolution tsunami modelling over
  cascadia.
\newblock {\em The Annals of Applied Statistics}, 12(4):2023--2053.

\bibitem[Higdon et~al., 2008]{higdon2008computer}
Higdon, D., Gattiker, J., Williams, B., and Rightley, M. (2008).
\newblock Computer model calibration using high-dimensional output.
\newblock {\em Journal of the American Statistical Association},
  103(482):570--583.

\bibitem[Hill, 1998]{hill1998mathematical}
Hill, R. (1998).
\newblock {\em The Mathematical Theory of Plasticity}, volume~11.
\newblock Oxford University Press.

\bibitem[Hockaday et~al., 2012]{hockaday2012rapid}
Hockaday, L., Kang, K., Colangelo, N., Cheung, P., Duan, B., Malone, E., Wu,
  J., Girardi, L., Bonassar, L., Lipson, H., Chu, C., and Butcher, J. (2012).
\newblock Rapid {3D} printing of anatomically accurate and mechanically
  heterogeneous aortic valve hydrogel scaffolds.
\newblock {\em Biofabrication}, 4(3):035005.

\bibitem[Ishwaran and Rao, 2005]{ishwaran2005spike}
Ishwaran, H. and Rao, J.~S. (2005).
\newblock Spike and slab variable selection: frequentist and bayesian
  strategies.
\newblock {\em The Annals of Statistics}, 33(2):730--773.

\bibitem[Joseph et~al., 2015]{joseph2015maximum}
Joseph, V.~R., Gul, E., and Ba, S. (2015).
\newblock Maximum projection designs for computer experiments.
\newblock {\em Biometrika}, 102(2):371--380.

\bibitem[Liu and Nocedal, 1989]{liu1989limited}
Liu, D.~C. and Nocedal, J. (1989).
\newblock On the limited memory {BFGS} method for large scale optimization.
\newblock {\em Mathematical Programming}, 45(1-3):503--528.

\bibitem[Liu and Guillas, 2017]{liu2017dimension}
Liu, X. and Guillas, S. (2017).
\newblock Dimension reduction for {Gaussian} process emulation: An application
  to the influence of bathymetry on tsunami heights.
\newblock {\em SIAM/ASA Journal on Uncertainty Quantification}, 5(1):787--812.

\bibitem[Mak et~al., 2018]{mak2018efficient}
Mak, S., Sung, C.-L., Wang, X., Yeh, S.~T., Chang, Y.~H., Joseph, V.~R., Yang,
  V., and Wu, C. F.~J. (2018).
\newblock An efficient surrogate model for emulation and physics extraction of
  large eddy simulations.
\newblock {\em Journal of the American Statistical Association},
  113(524):1443--1456.

\bibitem[Malfait and Ramsay, 2003]{malfait2003historical}
Malfait, N. and Ramsay, J.~O. (2003).
\newblock The historical functional linear model.
\newblock {\em Canadian Journal of Statistics}, 31(2):115--128.

\bibitem[Malvern, 1969]{malvern1969introduction}
Malvern, L.~E. (1969).
\newblock {\em Introduction to the Mechanics of a Continuous Medium}.
\newblock Number Monograph.

\bibitem[Martin et~al., 1998]{martin1998skeletal}
Martin, R.~B., Burr, D.~B., and Sharkey, N.~A. (1998).
\newblock {\em Skeletal Tissue Mechanics}, volume 190.
\newblock Springer.

\bibitem[Matheron, 1963]{matheron1963principles}
Matheron, G. (1963).
\newblock Principles of geostatistics.
\newblock {\em Economic Geology}, 58(8):1246--1266.

\bibitem[Mohammadi et~al., 2019]{mohammadi2019emulating}
Mohammadi, H., Challenor, P., and Goodfellow, M. (2019).
\newblock Emulating dynamic non-linear simulators using {Gaussian} processes.
\newblock {\em Computational Statistics \& Data Analysis}, 139:178--196.

\bibitem[Morris, 2012]{morris2012gaussian}
Morris, M.~D. (2012).
\newblock Gaussian surrogates for computer models with time-varying inputs and
  outputs.
\newblock {\em Technometrics}, 54(1):42--50.

\bibitem[Muehlenstaedt et~al., 2017]{muehlenstaedt2017computer}
Muehlenstaedt, T., Fruth, J., and Roustant, O. (2017).
\newblock Computer experiments with functional inputs and scalar outputs by a
  norm-based approach.
\newblock {\em Statistics and Computing}, 27(4):1083--1097.

\bibitem[Park and Casella, 2008]{park2008bayesian}
Park, T. and Casella, G. (2008).
\newblock The {Bayesian} lasso.
\newblock {\em Journal of the American Statistical Association},
  103(482):681--686.

\bibitem[Qian et~al., 2017]{qian2017quantitative}
Qian, Z., Wang, K., Liu, S., Zhou, X., Rajagopal, V., Meduri, C., Kauten,
  J.~R., Chang, Y.~H., Wu, C., Zhang, C., Wang, B., and Vannan, M.~A. (2017).
\newblock Quantitative prediction of paravalvular leak in transcatheter aortic
  valve replacement based on tissue-mimicking 3d printing.
\newblock {\em JACC: Cardiovascular Imaging}, 10(7):719--731.

\bibitem[Raghavan et~al., 1996]{raghavan1996ex}
Raghavan, M.~L., Webster, M.~W., and Vorp, D.~A. (1996).
\newblock Ex vivo biomechanical behavior of abdominal aortic aneurysm:
  assessment using a new mathematical model.
\newblock {\em Annals of Biomedical Engineering}, 24(5):573--582.

\bibitem[Ramsay, 2005]{ramsay2005functional}
Ramsay, J. (2005).
\newblock {\em Functional Data Analysis}.
\newblock Springer Series in Statistics, NY.

\bibitem[Rengier et~al., 2010]{rengier20103d}
Rengier, F., Mehndiratta, A., Von Tengg-Kobligk, H., Zechmann, C.~M.,
  Unterhinninghofen, R., Kauczor, H.-U., and Giesel, F.~L. (2010).
\newblock {3D} printing based on imaging data: review of medical applications.
\newblock {\em International Journal of Computer Assisted Radiology and
  Surgery}, 5(4):335--341.

\bibitem[Santner et~al., 2013]{santner2013design}
Santner, T.~J., Williams, B.~J., and Notz, W.~I. (2013).
\newblock {\em The Design and Analysis of Computer Experiments}.
\newblock Springer Science \& Business Media.

\bibitem[Sobol', 1967]{sobol1967distribution}
Sobol', I.~M. (1967).
\newblock On the distribution of points in a cube and the approximate
  evaluation of integrals.
\newblock {\em Zhurnal Vychislitel'noi Matematiki i Matematicheskoi Fiziki},
  7(4):784--802.

\bibitem[Sorensen et~al., 1987]{sorensen1987real}
Sorensen, H.~V., Jones, D., Heideman, M., and Burrus, C. (1987).
\newblock Real-valued fast {Fourier} transform algorithms.
\newblock {\em IEEE Transactions on Acoustics, Speech, and Signal Processing},
  35(6):849--863.

\bibitem[Stein and Corsten, 1991]{stein1991universal}
Stein, A. and Corsten, L. (1991).
\newblock Universal kriging and cokriging as a regression procedure.
\newblock {\em Biometrics}, 47(2):575--587.

\bibitem[Tseng, 2001]{tseng2001convergence}
Tseng, P. (2001).
\newblock Convergence of a block coordinate descent method for
  nondifferentiable minimization.
\newblock {\em Journal of Optimization Theory and Applications},
  109(3):475--494.

\bibitem[Wang and Xu, 2019]{wang2019gaussian}
Wang, B. and Xu, A. (2019).
\newblock Gaussian process methods for nonparametric functional regression with
  mixed predictors.
\newblock {\em Computational Statistics \& Data Analysis}, 131:80--90.

\bibitem[Wang, 2012]{wang2012bayesian}
Wang, H. (2012).
\newblock Bayesian graphical lasso models and efficient posterior computation.
\newblock {\em Bayesian Analysis}, 7(4):867--886.

\bibitem[Wang et~al., 2016]{wang2016controlling}
Wang, K., Zhao, Y., Chang, Y.-H., Qian, Z., Zhang, C., Wang, B., Vannan, M.~A.,
  and Wang, M.-J. (2016).
\newblock Controlling the mechanical behavior of dual-material {3D} printed
  meta-materials for patient-specific tissue-mimicking phantoms.
\newblock {\em Materials \& Design}, 90:704--712.

\bibitem[Wang and Berger, 2016]{wang2016estimating}
Wang, X. and Berger, J.~O. (2016).
\newblock Estimating shape constrained functions using gaussian processes.
\newblock {\em SIAM/ASA Journal on Uncertainty Quantification}, 4(1):1--25.

\bibitem[Zienkiewicz et~al., 1977]{zienkiewicz1977finite}
Zienkiewicz, O.~C., Taylor, R.~L., Zienkiewicz, O.~C., and Taylor, R.~L.
  (1977).
\newblock {\em The Finite Element Method}, volume~36.
\newblock McGraw-Hill London.

\end{thebibliography}

\normalsize

\newpage
\appendices

\numberwithin{equation}{section}

\section{Proof of Theorem 1}

For any $n \in \mathbb{N}, c_1, \cdots, c_n \in \mathbb{R}$ and function $I_1(\cdot),\cdots, I_n(\cdot) \in \mathcal{I}$, we have 
\begin{align}
\label{equ:111}
&\sum_{i=1}^n \sum_{j=1}^n c_i c_j \rho (I_i(\cdot),I_j(\cdot)) \\
=&\sum_{i=1}^n \sum_{j=1}^n c_i c_j \exp \left( - \int \theta(\xi)\left( \left|\hat{I}_i(\xi)\right|-\left|\hat{I}_j(\xi)\right|\right)^2 d\xi \right). 
\end{align}

Note that the Fourier transform $\mathcal{F}: \mathcal{I} \mapsto \mathcal{I} $, where $\mathcal{I}$ is the space of integrable functions $I(\cdot): \mathbb{R} \mapsto \mathbb{C}$, has a unique inverse $\mathcal{F}^{-1}: \mathcal{I} \mapsto \mathcal{I}$. Denote the standard Gaussian kernel as $K(\cdot,\cdot): |\mathcal{I}| \times |\mathcal{I}| \mapsto \mathbb{R}$,
\begin{equation}
K\left(F_1(\cdot),F_2(\cdot)\right)= \exp\left( -\int \theta(t) ( F_1(t)- F_2(t))^2 dt \right),
\end{equation}
where $|\mathcal{I}|$ is the space of integrable functions $F(\cdot): \mathbb{R}\mapsto\mathbb{R}$.
Since $K(\cdot,\cdot)$ is a positive definite kernel, for the selected $n$ and $c_1, \cdots, c_n$ in Equation (\ref{equ:111}), and any function $F_1(\cdot),\cdots, F_n(\cdot) \in \mathcal{I}$, we have,  
\begin{equation}
\sum_{i=1}^n \sum_{j=1}^n c_i c_j K (F_i(\cdot),F_j(\cdot)) \geq 0.
\end{equation}
Now let $F_k(\cdot)=\left|\hat{I}_k(\cdot)\right|$, where $k=1,2,\cdots,n$. This is possible because $\hat{I}_k(\cdot) \in \mathcal{I}$ and therefore $\left|\hat{I}_k(\cdot)\right| \in \mathcal{I}$. Thus, we have
\begin{equation}
\sum_{i=1}^n \sum_{j=1}^n c_i c_j K \left(\left|\hat{I}_i(\cdot)\right|,\left|\hat{I}_j(\cdot)\right|\right) =  \sum_{i=1}^n \sum_{j=1}^n c_i c_j \exp \left( - \int \theta(\xi)\left( \left|\hat{I}_i(\xi)\right|-\left|\hat{I}_j(\xi)\right|\right)^2 d\xi \right)  \geq 0
\end{equation}
In other words,
\begin{equation}
\sum_{i=1}^n \sum_{j=1}^n c_i c_j \rho (I_i(\cdot),I_j(\cdot)) = \sum_{i=1}^n \sum_{j=1}^n c_i c_j K (F_i(\cdot),F_j(\cdot)) \geq 0,
\end{equation}
i.e., the proposed SpeD correlation function is positive semi-definite.

\end{document}